\documentclass[journal,twoside,final]{IEEEtran}
\usepackage{amsmath}
\usepackage{cases}
\usepackage{amssymb}
\usepackage{amsthm} 
 \usepackage{multirow}
\usepackage{stfloats}
 \usepackage{array}  
\usepackage{booktabs}  
\usepackage{multicol}
\usepackage{bm}
\usepackage{bbm}
\usepackage{algorithmic}
\usepackage{algorithm}
 
\usepackage{graphicx}
\usepackage{epstopdf}
\usepackage{subfigure}
\usepackage{comment}
\usepackage{verbatim}
\usepackage{cite}
\usepackage{color}
\usepackage{amssymb}
\usepackage{mathrsfs}
\usepackage{amsfonts,amssymb}

\newtheorem{remark}{Remark}
\usepackage{graphicx}
\usepackage{float}
\usepackage{subfigure} 
\usepackage{stfloats}
\allowdisplaybreaks[4]

\begin{document}

\title{Fully Decentralized Federated Learning Based   Beamforming Design for UAV Communications}

\author{Yue~Xiao,     
          Yu~Ye,
          Shaocheng~Huang,
          Li~Hao,~\IEEEmembership{Member,~IEEE,} 
          Zheng~Ma,~\IEEEmembership{Member,~IEEE,} \\
          Ming~Xiao,~\IEEEmembership{Senior Member,~IEEE,
          Shahid~Mumtaz,~\IEEEmembership{Senior Member,~IEEE}    }         
      \thanks{Y. Xiao and L. Hao  are with the
School of Information Science and Technology, Southwest Jiaotong University, Chengdu 610031, China (e-mail: alice\_xiaoyue@hotmail.com;
lhao@home.swjtu.edu.cn).

Y. Ye, S. Huang, Z. Ma and M. Xiao are with the Division of Information Science
and Engineering, KTH Royal Institute of Technology, Stockholm, Sweden
(e-mail: {yu9, shahua, zma, mingx}@kth.se).

S. Mumtaz is with Instituto de Telecomunicacoes, Universidade de Aveiro, Campus Universitario de Santiago, 3810-193 Aveiro, Portugal (e-mail:smumtaz@av.it.pt).
}   
 \vspace{-0.2in}} 

\maketitle
 
\begin{abstract} To handle the data explosion in the era of internet of things (IoT), it is of interest to investigate the decentralized network, with the aim at relaxing the burden to central server along with keeping data privacy.
In this work, we develop a fully decentralized 
federated learning (FL) framework with an inexact stochastic parallel random  walk  alternating  direction method of multipliers (ISPW-ADMM). Performing more communication efficient and enhanced privacy preservation compared with the current state-of-the-art, the proposed ISPW-ADMM can be partially immune to the impacts from time-varying dynamic network and stochastic data collection, while still in fast convergence. Benefits from the stochastic gradients and biased first-order moment estimation, the proposed framework can be applied to any decentralized FL tasks over time-varying graphs. Thus to further demonstrate the practicability of such framework in providing fast convergence, high communication efficiency, and system robustness, we study the extreme learning machine(ELM)-based FL model for robust beamforming (BF) design in UAV communications, as verified by the numerical simulations.


\end{abstract}

%
\IEEEpeerreviewmaketitle
 
\section{Introduction}
Recently, the proliferation of Internet of Things (IoT) has triggered a surge in data traffic for future wireless networks. To alleviate such traffic conflicts in existing terrestrial infrastructures as well as provide cloud functionalities on demand, multi-dimensional integrated networking has been envisioned as the inevitable network architecture along with achieving the worldwide connectivity and coverage \cite{D20Nat}.
Fueled by such big data driven scenario and increasing computing power, 
{}
the  machine learning (ML)-enabled method is appealing in providing low computational cost and extrapolating new features
from environments \cite{LBH15DL}. 
However, the stringent requirement of stable/continuous network connections and substantial energy consumed by central controller pose rigorous challenges to the centralized scene composed of amount intelligent mobile agents (e.g., unmanned aerial vehicle (UAV)) \cite{ZS20arXiv}.
To this direction,
by decentralizing central service and spreading its burden to edge devices, the data computation and model training can be dealt locally in real-time.
Additionally, powered by the decentralized data management mechanism, general regulations governing data privacy can be satisfied \cite{YY20privacy}.

\begin{figure} [t]
\centering
 \includegraphics[width=80mm]{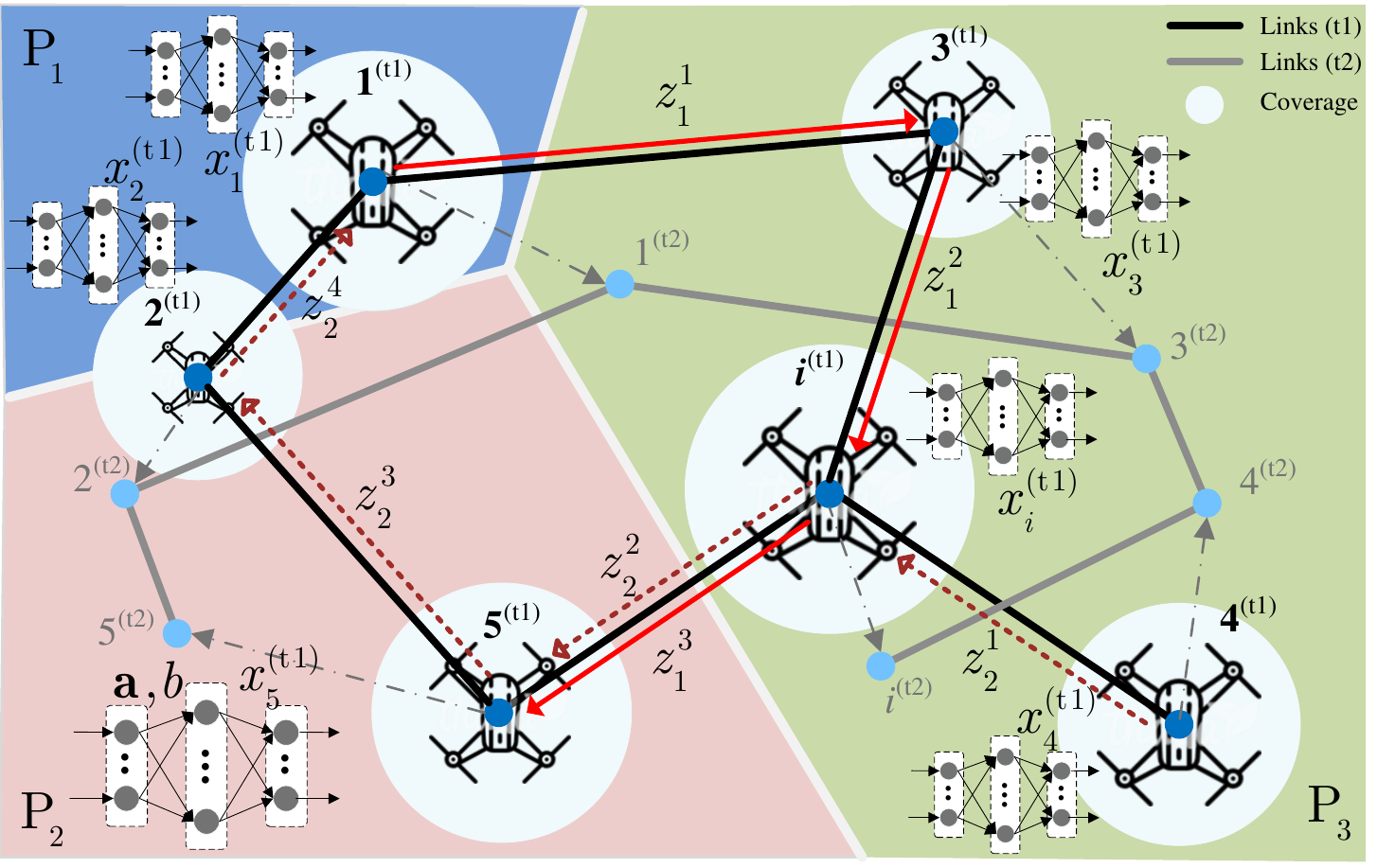}
\caption{Fully decentralized federated learning-assisted UAV communications.}
\label{[system model]}
\vskip -0.1 in
\end{figure}

In the context of decentralized manner,  federated learning (FL) has been recognized as an emerging approach to the collaborative model training via the topology of connected agents, while keeping the raw data locally dispersed\cite{K19Federated}. Towards this, both the privacy-preservation and communication/computation efficiency can be  guaranteed by leveraging a fully decentralized FL framework. 
Furthermore, to deploy production of such fully decentralized FL system in practice, the fleet of agents are expected to be capable of holding the reliable peer-to-peer communication,
 which is the key enabler for employing FL-based mechanism \cite{K19Federated}. For instance, referring to the 5G enhancement for UAV connections addressed in the latest 3GPP Release 17 \cite{20203GPPUAV} and the Flying Ad-Hoc Networks (FANETs) considered in IEEE 802.11 \cite{GJ15Survey}, the dynamic aerial network topology of UAV swarms can be built up to realize inter-node communications, which provide the suitability of investigating the FL-enabled UAV community \cite{ZS20arXiv}.

Aiming at mitigating the impact of dynamic communication environment on the reliability concerned in production system, the exploration of FL-enabled  networks has sparked an extreme interest to realize the ultra-reliable low-latency communication (URLLC) \cite{SBSD20TCOMM}, while the comprehensive research is still in infancy.
 In \cite{EC20arxiv}, neural network (NN) model employed at the base station (BS) is trained by gradient data collected from multiple users. However, the FL-based training process in \cite{EC20arxiv} is orchestrated by a central server, which implies the fragile state with a single point (i.e., central server) of failure.
The implementation of FL-based tasks optimization in wireless sensor networks are increasing in popularity for multiple local models \cite{ZS20arXiv}, \cite{L20arXiv}. The first work that applies FL scheme in UAV swarms can refer to \cite{ZS20arXiv}, in which the joint power allocation and flying trajectory design of UAV swarms are provided.  Moreover, a multi-dimensional contract-matching incentive mechanism for UAVs is designed by adopting FL-based sensing and collaborative learning scheme \cite{L20arXiv}. 
Practice wisdom encourages the application of theoretical research in real-world, yet current FL-based researches seldom consider the real-time stochastic data collection and time varying network topology.

In this work, we propose an inexact  stochastic parallel random walk alternating direction method of multipliers (ISPW-ADMM) algorithm that copes with decentralized FL tasks, along with maintaining the high communication/learning efficiency as well realizing enhanced privacy preservation. Besides, the proposed framework can also meet the challenges in time-varying connectivity graphs and stochastic data collection with potentially fast convergence. 
For a specific on-board mission in practice, the robust beamforming (BF) design is first realized by adopting local extreme learning machine (ELM) model. Then, all local models are supposed to reach the global consensus solution, by integrating the decentralized FL framework. Through numerical results, the proposed algorithm is validated to be both communication and time efficient.

\section{Fully Decentralized Federated Learning Framework} \label{[System Framework]}
 
\subsection{Fully Decentralized Framework}\label{Fully_Decentralized_Framework}
As illustrated in Fig. \ref{[system model]}, a swarm of traveling UAVs (agents) $(\mathcal{N} =\{1,...,N \})$ are considered to provide wireless services to the ground terminals in geographically distributed regions.
With the aim to solve the decentralized consensus optimization problem in such multi-agent system \cite{MY19IOT}, we have  
\begin{equation} \label{Centralized_Consecus}
\min _{x\in { {\mathbb{R}}^{p}}}~\sum _{i=1}^{N}\mathbb{E}_{\bm \zeta_i}[f_{i}(x_i;\bm \zeta_i)],~s.t.~x_i=z,\forall i\in\mathcal{N},  
\end{equation}
with $f_i: \mathbb{R}^{p} \rightarrow \mathbb{R} $ being the local loss function of model weights $x_i\in\mathbb{R}^p$ privately held by UAV $i\in\mathcal{N}$. In FL tasks \cite{K19Federated}, the agents cooperate with each other to find a global model (i.e., $z$) through drawing sequences of identical and independent (i.i.d.) observations from the random vector $\{\bm \zeta_i|i\in\mathcal{N} \}$. Specifically, $\bm\zeta_i$ obeys a fixed distribution $\text{P}_l$ out of the set $\textbf{P}=\{\text{P}_1,...,\text{P}_L| L\in\mathbb{N}^+\}$. 
Hereinafter, we denote $\mathbb{E}_{\bm\zeta_i}[f_i(x_i;\bm\zeta_i)]=f_i(x_i)$ for simplicity.
Refer to the fully decentralized manner presented as PW-ADMM in \cite{YY20CL}, problem (\ref{Centralized_Consecus}) can be equivalently rewritten as
\vspace{-0.05in}
\begin{equation} \label{Decentralized_Consecus}
\begin{aligned}
 \min _{\boldsymbol {x}, \boldsymbol {z}}~\sum _{i=1}^{N}f_{i}(x_{i}), ~\quad s.t.~ {{\mathbbm {1}}_p}\otimes \frac {1}{M}\sum _{m=1}^{M}z_{m}-\boldsymbol {x}=\bm{0},
\end{aligned}
\end{equation}
with $\otimes$ denoting Kronecker product, ${\mathbbm {1}}_p= [1, \cdots, 1]^T \in  \mathbb{R}^{p}$ and $\boldsymbol{x}=[x_1,\cdots, x_N]\in \mathbb{R}^{pN}$.  $\boldsymbol{z}=[z_1,\cdots, z_M]\in \mathbb{R}^{pM}$ denotes the tokens held by random walks $\mathcal{M}=\{1,...,M \}$. For FL in decentralized manner, the global consensus solution shall be the  average value of all the tokens, i.e.,  $\frac{1}{M}\sum_{m=1}^Mz_m$. Thus, the augmented Lagrangian for (\ref{Decentralized_Consecus}) is given by
\begin{equation}
\begin{aligned}
 \hspace{-0.5pc}\mathcal {L}_{\rho }(\boldsymbol {x},\boldsymbol {z},\boldsymbol {y})= \sum _{i=1}^{N}f_{i}(x_{i})+& \bigg\langle \boldsymbol {y }, {{\mathbbm {1}}_p}\otimes \frac {1}{M}\sum _{m=1}^{M}z_{m}-\boldsymbol {x} \bigg\rangle \\       +  & \frac {\rho }{2} \bigg\Vert {{\mathbbm {1}}_p}\otimes \frac {1}{M}\sum _{m=1}^{M}z_{m}-\boldsymbol {x} \bigg\Vert^{2},    
\end{aligned}  
\end{equation}  
where $ \boldsymbol {y }$ is a Lagrange multiplier and $\rho>0$ denotes the constant. Following updates of synchronous inexact ADMM \cite{chang2014multi} and the proximal stochastic ADMM \cite{huang2018mini}, the solution to (\ref{Decentralized_Consecus}) can be obtained in iterations, wherein the updates for the $(k+1)$-th iteration follow
 \begin{subequations}
 \begin{align} 
&x_i^{k+1}:=\left\{\begin{aligned}
&\arg \min_{x_i} \hat{\mathcal{L}}_{\rho,i  }^{k} (x_i, \bm{z}^{k } , y_i^k ) ,~i=i_{m_{k+1}} ;\\
&x_i^{k },~\text{otherwise};
\end{aligned}  \right. \label{xup1}\\
&y_i^{k+1}:=\left\{\begin{aligned}
& y_i^{k } + \gamma\rho \bigg(\frac{1}{M}\sum_{m=1}^{M}z_m^{k }-x_{i}^{k +1} \bigg),~i=i_{m_{k+1}};\\
& y_i^{k },~\text{otherwise};
\end{aligned}  \right. \label{yup1} \\
 &z_m^{k+1}:=\left\{\begin{aligned} \label{zup1}
 &\arg \min_{z_m} \mathcal{L}_{\rho} (\bm{x}^{k+1},z_m, \bm{z}_{-m}^k  ,\bm{y}^{k+1} ),~m=m_{k+1} ;\\
 &z_m^{k}, \text{otherwise}; 
 \end{aligned}  \right.   
 \end{align}	
  \end{subequations}
where $\bm{z}_{-m}^k=\{z_1^k,...,z_{m-1}^k,z_{m+1}^k,...,z_M^k   \} $, and
\begin{equation}
\begin{aligned}
 &\hat{\mathcal{L}}_{\rho,i }^{k} (x_i, \bm{z}^{k } , y_i^k )= g_i(x_i^k;\bm{\zeta}_i^k)(x_i-x_i^k)  \\
 &~~~~~~~~+ \frac {\rho }{2} \bigg\| \frac {1}{M}\sum _{m=1}^{M}z_{m}^k-x_i +\frac{y_i^k}{\rho} \bigg\|^{2}+\frac{\tau}{2}\| x_i - x_i^k\|^2,    
\end{aligned}
\end{equation}
$\tau$ and $\gamma$ are step sizes for primal and dual updates, respectively, while $g_i(x_i; \bm{\zeta}_i)\triangleq \nabla f_i(x_i;\bm{\zeta}_i)$ is the stochastic gradient. According to \cite{YM19arXiv}, the convergence speed of ADMM with first-order approximation may degrade from traditional ADMM. Besides, the stochastic property of $g_i(\cdot)$ will introduce data variance in the primal update \cite{zheng2016fast}. To this point, the biased first-order moment estimate is further proposed to stabilize and speed up convergence for stochastic updates, that is 
  \begin{equation}\label{fbe}
		    \mu_{i}^{k+1}:= \eta \mu_{i}^{k} + (1-\eta) g_{i} (x_{i}^k;\bm{\zeta}_{i}^k ),~i=i_{m_{k+1}},
 \end{equation}
where $\eta\in[0,1)$ denotes exponential decay rates for the first order moment estimate.
Till now, the updates of tokens $\bm z$ are still in centralized manner.  

Then by initializing $\bm{z}^0= \bm{x}^0= \bm{y}^0=\bm{0}$ in (\ref{zup1}), it is clear to find that $z_{m}^{k+1}$ can be \textit{incrementally} updated as
\begin{equation}\label{zup}
\begin{aligned}
	z_{m}^{k+2} = z_{m }^{k+1} + \frac{M}{N} \bigg[\bigg(x_i^{k+1}-\frac{y_i^{k+1}}{\rho} \bigg) - \bigg(x_i^{k }-\frac{y_i^{k }}{\rho} \bigg)\bigg], 
	\end{aligned}
\end{equation} 
  where $m=m_{k+1}$ and $i=i_{m_{k+1}}$ denote the activated random walk and agent, respectively. That is, the update of $z_m$ does not require the information from other tokens, i.e., $\bm{z}_{-m}$. Thus, the updates can be carried out in  parallel eventually. Following conventional PW-ADMM \cite{YY20CL}, the updates given by (\ref{xup1})-(\ref{zup1})  can be explained in asynchronous manner, by approximating $\frac{1}{M}\sum_{m=1}^M z_m$ with received token $z_m$ in (\ref{xup1}) and (\ref{yup1}). Since all agents and parallel random walks can keep independent updating clock, two variables are introduced, i.e., $k_i$ and $s_m$ corresponding to agent $i\in\mathcal{N}$ and random walk $m\in\mathcal{M}$, respectively.

  Different from the conventional PW-ADMM, the mobility of such UAV swarm system indicates the time-varying undirected graph $\mathcal{G}^{(t)}=( \mathcal{E}^{(t)},\mathcal{N} )$, where $t$ is time stamp and $\mathcal{E}^{(t)}$ is the set of links at time $t$. 
  In specific, if agent $j$ travels within the communication range of $i$ at time $t$,  we have $(i,j)\in\mathcal{E}^{(t)}$.
  By defining $\overline{\mathcal{N}}_i^{(t)}$ as the set of neighboring agents for agent $i$ at time $t$, we summarize the inexact stochastic PW-ADMM (ISPW-ADMM) in Algorithm 1. Note that with $M=1$, the parallel random walk token transmission reduces to the conventional random walk strategy \cite{MYHY20TSP}.
     \begin{algorithm}[t]
	\caption{ISPW-ADMM } 
	\begin{algorithmic}[1]
		\STATE \textbf{initialize}:  $\{z^0 = x_i^0 = y_i^0= \mu_i^0=\bm{0}, k_i=0,\eta, \gamma |i\in\mathcal{N}\}$; 
	    \STATE \textbf{Algorithm for the $m$-th random Walk:}
		\FOR{$s_m=0,1,...$} 
		\STATE wait token $z^{s_m}_m$ arrive at agent $i=i_{s_m}$;
		\STATE draw i.i.d. samples $\bm{\zeta}_i^{k_i} \sim \overline{\text{P}}_{i}^{(t)}$ with $i=i_{s_m}$;
		\STATE update $\mu_{i}^{k_i+1}$ by (\ref{fbe}) with $i=i_{s_m}$; 	
		\STATE update $x_{i }^{k_i+1}$ by (\ref{xup}) with $i=i_{s_m}$;
		\begin{equation}\label{xup}
		\begin{aligned}
			    x_{i }^{k_i+1}&:=\arg\min_{x_{i}}  \mu_{i }^{k_i+1}  (x_{i }-x_{i }^{k_i} ) \\&~~~ +\frac{\rho}{2} \bigg\| z^{s_m}_m-x_{i} +\frac{y_{i}^{k_i} }{\rho} \bigg\|   + \frac{\tau}{2}\|x_{i}-x_{i}^{k_i} \|;
		\end{aligned}
		\end{equation}
		\STATE update $y_{i}^{k_i+1}$ by (\ref{yup}) with $i=i_{s_m}$;
		\begin{equation}\label{yup}
		    y_i^{k_i+1}:=y_i^{k_i} + \gamma \rho (z_m^{s_m} - x_i^{k_i+1});
		\end{equation}
		\STATE update $z^{s_m+1}_m$ according to (\ref{zup});
		\STATE set $k_i\gets k_i + 1$ with $i=i_{s_m}$; 		
		\STATE choose $i_{s_m+1}(\in \overline{\mathcal{N}}_{i}^{(t)}  )$ according to $\bm{P}^{(t)}_{i}$ with $i=i_{s_m}$; 
		\STATE send token $z_m^{s_m+1} $ to agent $i_{s_m+1} $;  
		\ENDFOR 		
	\end{algorithmic} 
\end{algorithm}
In ISPW-ADMM, after the token $z_m^{s_m}$ arriving at agent $i=i_{s_m}$ via the $m$-th random walk, the collected samples $\bm \zeta_{i}^{k_i}\sim \overline{\text{P}}_i^{(t)}$ are used for training local model $x_{i}$, where $\overline{\text{P}}_i^{(t)}\in \textbf{P}$ is determined by the location of UAV $i$ at time $t$. One more, the transition of token $z_m^{s_m+1}$ follows the embedded Markov chain with time-varying probability matrix ${\bm{{P}}}_{i}^{(t)}$ \cite{YY20CL}.

\subsection{Discussions}
Regarding the integration of the fully decentralized framework with FL learning model,
 the local NN model (i.e., $x_i$) is first designed to output local solutions, then individual NN models gradually reach the desired global solution via the dynamic connectivity graph $\mathcal{G}^{(t)}$. 
\begin{remark}\label{Rm1}
The ISPW-ADMM can be applicable to any decentralized FL tasks over time-varying graphs.
\end{remark}
{}
In what follows, preliminary statements are entailed for meeting the challenges in the FL-based model with dynamic connections,
\begin{itemize}
\item[$-$]  {\textit{High communication/learning efficiency:} } The W-ADMM achieves the less communication cost with single random node being activated in sequence, while PW-ADMM allows multiple walks in parallel to reduce the running time. Hence according to \cite{YY20CL}, the proposed ISPW-ADMM can be utilized to trade-off the communication cost and running time.
\item[$-$]  {\textit{Enhanced privacy preservation:} } 
To further develop privacy preserving, partially homomorphic encryption \cite{Alex20encrop} can be exploited in the transition of tokens (i.e., $\bm z$) to protect the exchanged information between the connected agents .
\item[$-$]  {\textit{Time varying topology:} }  Apart from getting more relevant to practical mobile communications, the multiple random walk mechanism in ISPW-ADMM allows each node to be traversed equally in long-run updating, even with dynamic connected graph. By doing so, the time-varying matrices (i.e., ${\bm{{P}}}_{i}^{(t)}$ and $\mathcal{G}^{(t)}$) will not heavily hurt the resulting averaged global performance.  
\item[$-$]  {\textit{Stochastic database:} } 
In sight of the unbalanced/biased database collected by the traveling agents, the proposed scheme can be realized potentially converge fast by utilizing the stochastic gradients and biased estimation on first-order moment.

{}

\end{itemize}


\section{FL-based beamforming design}
Aiming at the concrete on-board mission encouraged by the proposed full decentralized FL framework, we first present local ELM model at single UAV agent for robust BF design with respect to noisy channel state information (CSI) in this section.
Inspired by the multiple random walk mechanism ISPW-ADMM as stated in Section  \ref{[System Framework]},  all local models can gradually converge to consensus. Eventually, the desired global BF design can be realized, while considering the dynamic UAV swarms and stochastic CSIs collection during the traveling.
{}
\subsection{Beamforming Design}
For UAV MIMO communications, the millimeter wave (mmWave) channel coefficient experienced from UAV  $i\in\mathcal{N}$ to the ground terminal is denoted by ${{\bf{H}}_{{{i}}}} \in {\mathbb{C}}^{N_r \times N_t}$, with $N_t$ and $N_r$ being the transmit and receive antennas equipped at UAV and the ground node, respectively.
Accordingly, the optimal fully digital (FD) beamforming ${\bf{F}}^{\text{opt}}$ shall be designed to maximize the  achievable rate obtained over the mmWave channel, that is, 
\begin{equation}\label{Rate}
    \begin{aligned}
{\bf{F}}^{\text{opt}}=\arg\max_{{\bf{F}}}  \log_2\left(\left\vert{\bf{I}}+{\rho_{{r}}} {\bf{H}}_i {\bf{F}} {\bf{F}}^{H} {\bf{H}}_i^{H} 
\right\vert\right),
 \end{aligned}
\end{equation}
where ${\rho_{{r}}}$ denotes the  average received SNR. Based on the singular value decomposition (SVD) of ${\bf{H}}_i$, (\ref{Rate}) can be reformulated as 
\begin{equation}\label{SVD_Rate}
    \begin{aligned}
{\bf{F}}^{\text{opt}}=\arg\max_{{\bf{F}}}  \log_2\left(\left\vert{\bf{I}}+{\rho_{\rm{r}}} {\mathbf{\Sigma}}^2  {\mathbf{V}}^{H} {\bf{F}} {\bf{F}}^{H}  
{\mathbf{V}} \right\vert\right),
 \end{aligned}
\end{equation}
where ${\mathbf{\Sigma}}$ and ${\mathbf{V}}$ denote  the diagonal matrix and right unitary matrix of ${\bf{H}}_i$ respectively, deriving from ${\bf{H}}_i={\bf{U}} {\bf{\Sigma}}  {\bf{V}}^{H}  $. For the BF design with $N_s$ transmitted data streams, the right unitary matrix shown in (\ref{SVD_Rate}) can be separated as ${\bf{V}}=\left[{\bf{V}}^{(1)}, {\bf{V}}^{(2)}\right]$, with ${\bf{V}}^{(1)}\in {\mathbb{C}}^{N_t \times N_s}$ and ${\bf{V}}^{(2)}\in {\mathbb{C}}^{N_t \times \text{rank}({\bf{H}}_i)-N_s}$.
Thus, one can find that the optimum FD beamformer can be simply expressed by 
\begin{equation}\label{F_opt}
    \begin{aligned}
{\bf{F}}^{\text{opt}}={\bf{V}}^{(1)},
 \end{aligned}
\end{equation}
under the approximation of that ${\bf{V}}^{(2)} {\bf{F}}^{\text{opt}} \approx \bf{0}$ \cite{AR14TWC}. Even for the hybrid analog and digital BF design, it's equivalent to approaching the performance of FD beamformer by minimizing the Frobenius norm of the gap between such two schemes, e.g., the orthogonal matching pursuit (OMP) algorithm \cite{AR14TWC}. 
However, as a matrix factorization technique, large matrices manipulation makes the SVD-related algorithm not fit in the main memory of mobility drones, i.e., the complexity for computing the SVD of a $N_r \times N_t$  matrix is $ \mathcal{O}\left(N_t N_r \
\min(N_t, N_r)\right)$ \cite{SVD17arxiv}. Towards this, it might not be flexible to apply traditional BF design in the fast random access communications by observing diverse CSIs in practice.
\subsection{FL-based Robust BF Design}
In order to circumvent the high complexity of optimization algorithms, it's imperative to build up an ML framework with low computational complexity, which is also capable of extrapolating new features from a limited set of noisy training data \cite{HYX20arXiv}.
For easier implementations, single layer feedforward neural network (SLFN) has demonstrated powerful potentials for data regression and classification in faster learning speed and least human intervene compared with the conventional ML technique. Furthermore, without need of tuning the hidden layer parameters, ELM has been developed for the ``generalized'' SLFN, performing in low complexity \cite{HG11}. Thus, due to the hardware constraint of  energy-limited devices (e.g., UAV), ELM  has been verified to be the fast technique for the BF design in low latency communication \cite{HYX20arXiv}. 
By deploying ELM model at each UAV, the output weights $\boldsymbol{x}_i$ of ELM scheme at UAV $i$ shall be learned form a training database $\bm\zeta_{i(l,t)}$. In detail, the term $i(l,t)$ is decided by the location of UAV $i$ traveling at time $t$, by which the data distribution follows the distribution $\text{P}_l$. For simplicity, we denote $i={i(l,t)}$.
Similarly, we have $\bm\zeta_i=\bm\zeta_{i(l,t)}$ hereinafter, given by
\begin{equation}\label{Training_data}
\begin{aligned}
\bm\zeta_i =\{(\boldsymbol{S}_{i,d},\boldsymbol{T}_{i,d}) \vert~ d=1,..., D_{i} \},
 \end{aligned}
\end{equation}
with $\boldsymbol{S}_{i,d}$ and $\boldsymbol{T}_{i,d}$ being the sample and target for the $d-$th training data fed into the ML model at UAV $i$, respectively. $D_i$ denotes the number of training samples collected by UAV $i$. More specific, the input format of the training samples to ELM model is given by
\begin{equation}\label{Training_data}
\begin{aligned}
\boldsymbol{S}_{i,d}=\left[{\textrm{Re}}\left({\textrm{vec}}({{\bf{H}}^{(r,c)}_{{{i}}}})\right),\textrm{Im}\left({\textrm{vec}}({{\bf{H}}^{(r,c)}_{{{i}}}})\right)\right]^{T} 
\in \mathbb{R}^{2 N_r N_t},
 \end{aligned}
\end{equation}
where $\textrm{Re}(\cdot)$ and $\textrm{Im}(\cdot)$ denote the real and imaginary part, respectively. Moreover, we have 
\begin{equation}
    \{{\bf{H}}^{(r,c)}_{{{i}}}=
\mathcal{CN}({{\bf{H}}^{(r)}_{{{i}}}}, \sigma^2_{\text{Train}})\vert r\in(1,...,R_i), c\in(1,..., C_i)\},
\end{equation}
which are the noisy channels based on $R_i$ different realizations. 
For each realization, we assume there are $C_i$ samples for training the ELM model. Till now,  all the collected samples at UAV $i$ is $D_i=R_i C_i$. To evaluate the variance of the white Gaussian noise (AWGN) added to desired signals, each channel entry can be explained by $\text{SNR}_{\text{Train}} (\text{dB})=\big|{[{\bf{H}}^{(r,c)}_{{{i}}}]_{(.,.)}}\big|^2-[\sigma^2_{{\text{Train}}{(.,.)}}]$, with $\sigma^2_{\text{Train}}$  being the variance of noise.
Similarly, the target can be obtained by substituting ${\bf{H}}^{(r)}_{{{i}}}$ into (\ref{F_opt}), that is 
\begin{equation}\label{Training_target}
\begin{aligned} 
\boldsymbol{T}_{i,d}=\left[\text{Re}\left(\text{vec}\big({\bf{F}}_{i}^{\text{opt}{(r)}}\big)\right),\text{Im}\left({\text{vec}}\big({\bf{F}}_{i}^{\text{opt}{(r)}}\big)\right)\right] 
\in \mathbb{R}^{2 N_t N_s}.
 \end{aligned}
\end{equation}
Thus, the accurate labels of the training data samples have been provided appropriately. 

Refer to the principles of ELM in \cite{HG11}, the output of an ELM model at UAV $i$ related to $Q$ hidden nodes is given by
\begin{equation}\label{ELM_output}
\begin{aligned}
\boldsymbol{Y}_{i,Q}(\boldsymbol{S}_{i,d})=\sum_{q=1}^{Q}x_{i,q} G_{i, q}(\boldsymbol{S}_{i,d})=\boldsymbol{G}_i(\boldsymbol{S}_{i,d})x_i,
 \end{aligned}
\end{equation}
 where $x_i\in \mathbb{R}^{Q\times2N_tN_s}$ is the output weight, $\boldsymbol{G}_i(\boldsymbol{S}_{i,d})$ denotes the feature mapping relation of the training input $\boldsymbol{S}_{i,d}$ from $2{N_r}{N_t}$ to
$Q$ dimensions. With the given randomized weights connecting the $q$-th hidden node and the input nodes $\{\bold{a}_{q}\}$ and the bias of $q$-th hidden node $\{b_{q}\}$, a nonlinear piecewise continuous function can be used at the hidden node as its activation function \cite{HG11}. For example, the well-known Sigmoid function is given by 
 \begin{equation}\label{ELM_output}
\begin{aligned}
G_{i, q}(\bold{a}_{q}, b_{q} \boldsymbol{S}_{i,d})
=(1+\exp(-(\bold{a}_{q}\boldsymbol{S}_{i,d}+b_{q})))^{-1}.
 \end{aligned}
\end{equation}
To this end, we implement ISPW-ADMM with the local loss function
\begin{equation}\label{optimum_LossF}
\begin{aligned}
 f_{i}(x_i;\bm \zeta_i  )= \frac{1}{2\lambda_{e}}\| \bm{W}_i  x_i- \bm{T}_{i}\|^{2} + \frac{1}{2}\| x_i\|^{2},
\end{aligned}
\end{equation}
with $\boldsymbol{W}_i=[\bm{G}_i(\bm{S}_{i,1}),...,\bm{G}_i(\boldsymbol{S}_{i,D_i})]^T\in \mathbb{R}^{Q\times2N_rN_t}$ being the hidden layer output matrix at agent $ i$.
$\lambda_e$ denotes the tradeoff parameter between separating margin and training error. 

 
\begin{figure*}[t]
\vskip -0.2 in
\minipage{0.3333\textwidth}
  \includegraphics[width=\linewidth]{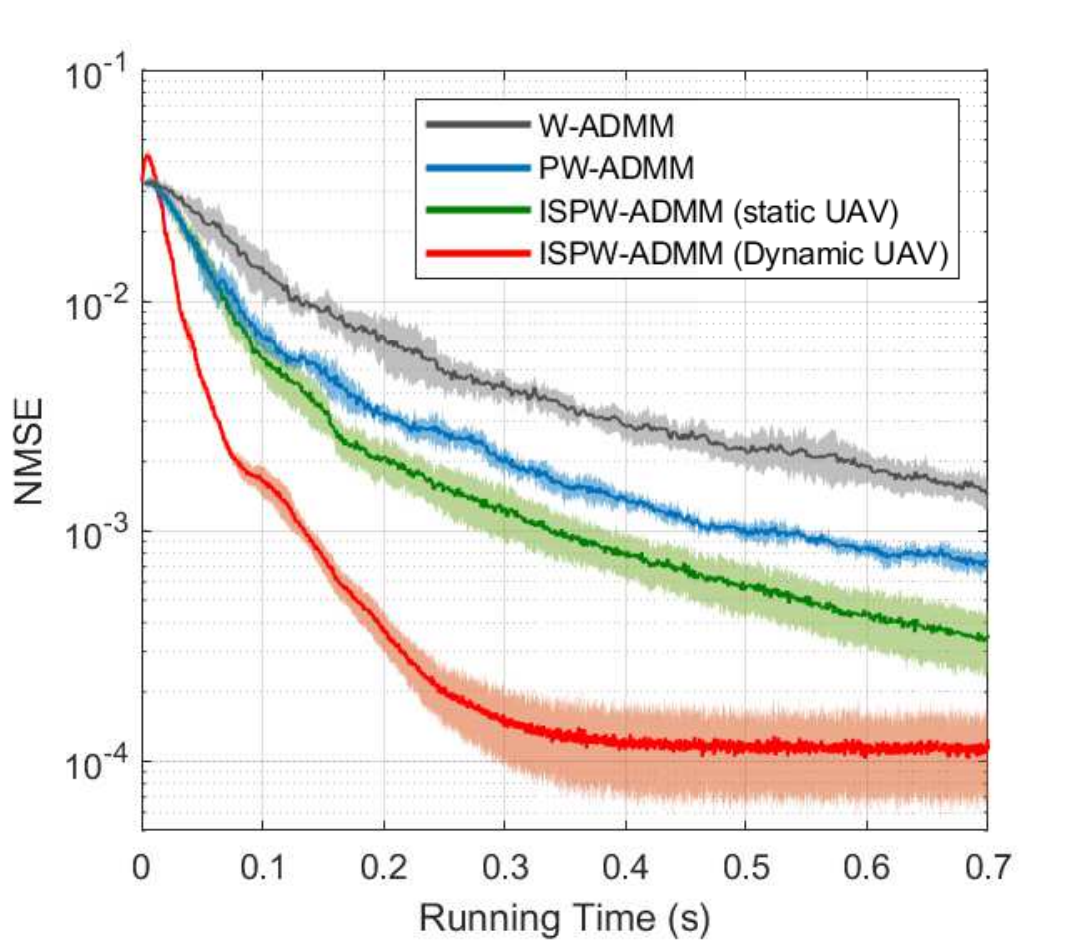}
  \caption{ Testing NMSE vs running time.}\label{[NMSE_Time]}
\endminipage\hfill
\minipage{0.3333\textwidth}
  \includegraphics[width=\linewidth]{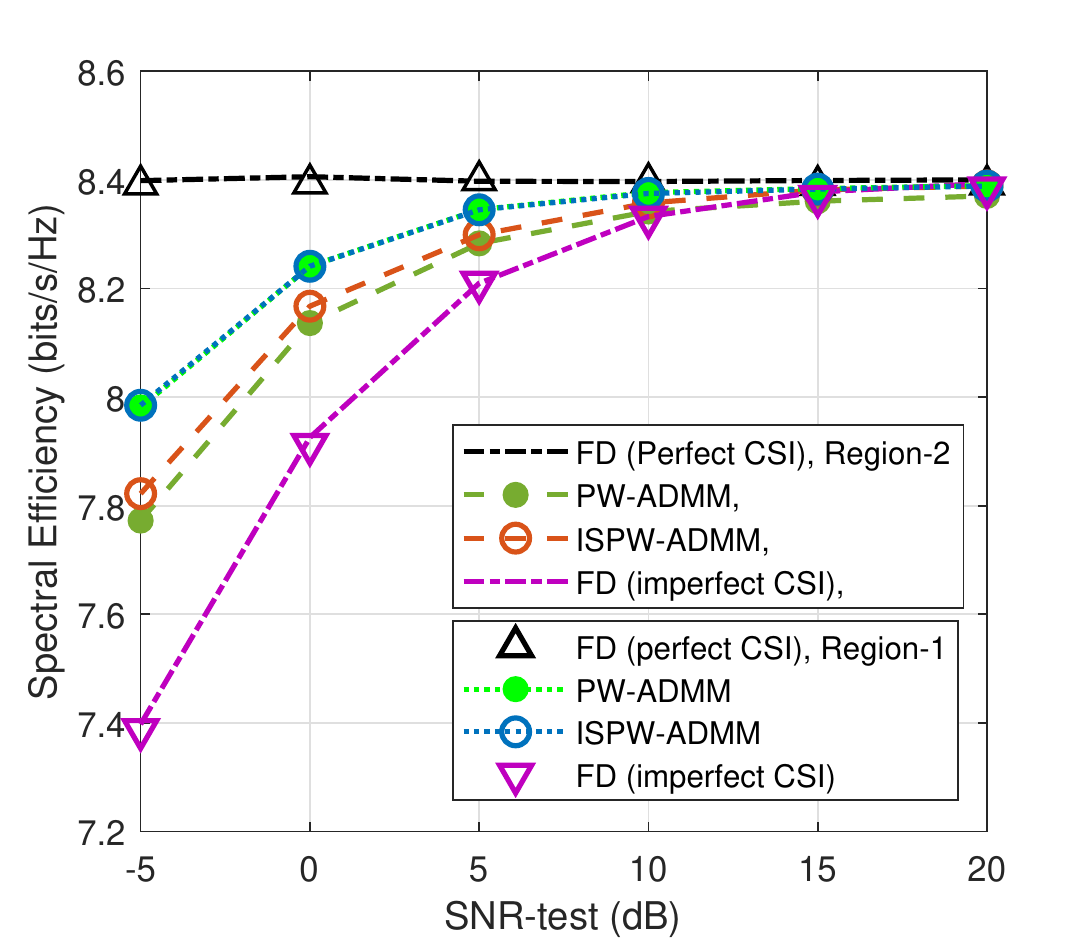}
  \caption{Testing spectral efficiency vs $\text{SNR}_{\text{Test}}$.}\label{[Rate_SNR]}
\endminipage\hfill
\minipage{0.3333\textwidth}%
  \includegraphics[width=\linewidth]{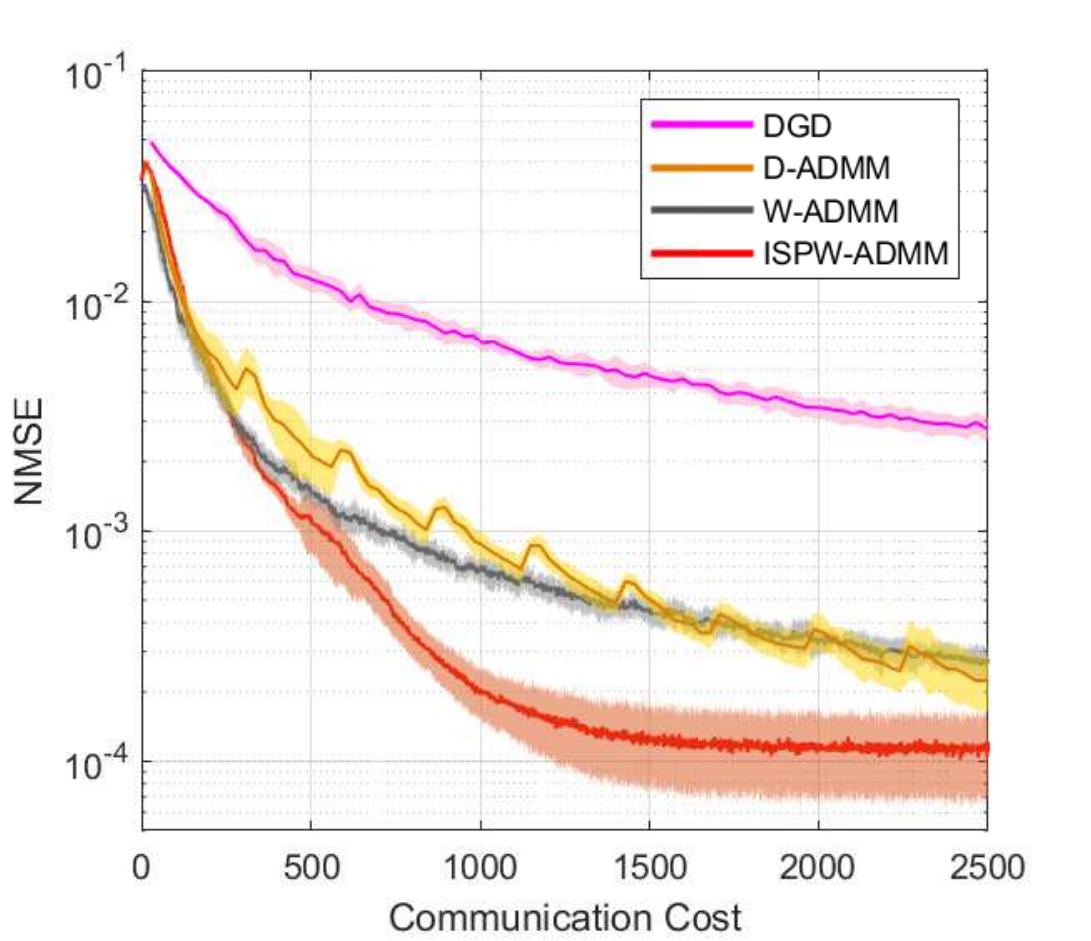}
  \caption{ Testing NMSE vs communication energy.}\label{[NMSE_Commu]}
\endminipage
\vskip -0.2 in
\end{figure*}
\section{Simulation}
{}
To verify the practicality  of ISPW-ADMM scheme coped with the energy effective and fast network access in UAV communications, we compare  ISPW-ADMM with state-of-the-art  decentralized optimization methods, 
including W-ADMM, PW-ADMM, and decentralized gradient descent (DGD) ($\alpha=10^{-2}$) \cite{DGD}, distributed-ADMM (D-ADMM) ($\rho=2$) \cite{d-admm}.  
If not otherwise specified, the values of all parameters are given by $N=10$, $M=2$, and $L=2$, $N_t=16$, $ N_r=1$, and $R_i=10$, $Q=200$,  and $\lambda_e=10^{-2}$, $\eta=0.95$, $\rho=2$,  and $\gamma=1$, $\tau=10$.
Without loss of generality, we assume that $\text{SNR}_{\text{Train}}=\text{SNR}_{\text{Test}}$, that is, the SNR utilized for data training and testing data are set the same.

In Fig. \ref{[NMSE_Time]}, the  
normalized mean square error (NMSE) performance of different decentralized algorithms are presented versus running time. It's intuitive to find that the proposed ISPW-ADMM-based training in the dynamic connectivity graph of UAV swarm is the most efficient in  time cost.  
Even for training by static connected UAV swarm, the proposed ISPW-ADMM scheme still performs better than the synchronous W-ADMM and PW-ADMM method. This is because ISPW-ADMM with inexact updates outperforms the conventional stochastic synchronous and asynchronous method. 



While considering the robust beamforming design in the proposed FL-enabled ELM learning model, the spectral efficiency achieved by the aforementioned decentralized methods are  presented versus different $\text{SNR}_{\text{Test}}$ in Fig. \ref{[Rate_SNR]}, with perfect CSI and imperfect CSI assumed, and we have $\rho_r=20$ dB. Overall, the performance of all schemes increase with enlarging $\text{SNR}_\text{Test}$, apart from the FD beamformer with perfect CSI assumed which is immune to any noise. The benchmark is presented by the imperfect CSI-based FD scheme without ELM training. Recall to that ELM works on extracting new features from noisy database, the ELM-based FD is robust against the noisy channels, even with extreme large noise added. Moreover, due to that the classes of channels in one region are less than those in multi-region (e.g., two regions), the case of one region (denoted by Region-1) performs better than that of two regions (i.e., Region-2). By the results, one can find that the stochastic data driven ISPW-ADMM algorithm is superior in achieving higher spectral efficiency, compared with PW-ADMM, especially as shown in the case of Region-2.

The testing NMSE over communication cost is shown in Fig. \ref{[NMSE_Commu]}, which is another vital metric concerned in the energy-limited UAV community, i.e., less communication cost indicates higher communication efficiency. Herein, we consider the unicast and the communication cost for transmitting a $Q$-dimensional vector is unit 1. It's clear to see that the ISPW-ADMM is the most energy efficient proposal compared with W-ADMM, DGD, and D-ADMM. This is due to that all the links in DGD and D-ADMM are active in each iteration, which consume more energy for information sharing. Thus, the proposed ISPW-ADMM is valid to be effectively used in realizing the robust beamforming design for fully decentralized UAV communications.

\section{Conclusion}
In this work, we have proposed the multiple random walk  mechanism for ISPW-ADMM based consensus optimization. 
By which, any fully decentralized FL tasks over time-varying graphs can be solved, along with maintaining high communication/learning efficiency and enhanced privacy preservation. Moreover, with the unbalanced data collected in practice, the stochastic gradients and biased first-order moment estimation leveraged in ISPW-ADMM can guarantee the fast convergence. Then, a specific on-board mission is presented to further verify the effectiveness of ISPW-ADMM in wireless applications, i.e., the ELM-enabled robust beamforming design, as it is verified by the presented numerical results.

\bibliographystyle{IEEEtran}
\bibliography{XY_UAV_FL}

\end{document}